\documentclass[final,5p,times,twocolumn]{elsarticle}
\usepackage{times,amsmath,amssymb,amsfonts,graphicx,bbm}
\usepackage{lineno}
\newcommand{\Tr}{\mathop{\text{Tr}}\nolimits}
\journal{Physica A}
\biboptions{square}
\begin{document}
\begin{frontmatter}
\title{Quantum probes for fractional Gaussian processes}
\author{Matteo G. A. Paris}\ead{matteo.paris@fisica.unimi.it}
\address{Dipartimento di Fisica dell'Universit\`a 
degli Studi di Milano, I-20133 Milano, Italy}
\date{\today}
\begin{abstract}
We address the characterization of classical fractional random noise via quantum
probes. In particular, we focus on estimation and discrimination
problems involving the fractal dimension of the trajectories of a system
subject to fractional Brownian noise.  We assume that the classical
degree of freedom exposed to the environmental noise is coupled to a
quantum degree of freedom of the same system, e.g. its spin, and exploit
quantum limited measurements on the spin part to characterize the
classical fractional noise. 
More generally, our approach may be applied to any two-level system
subject to dephasing perturbations described by fractional Brownian 
noise, in order to assess the precision of quantum limited measurements 
in the characterization of the external noise.
In order to assess the performances of
quantum probes we evaluate the Bures metric, as well as the Helstrom and
the Chernoff bound, and optimize their values over the interaction time.
We find that quantum probes may be successfully employed to obtain a
reliable characterization of fractional Gaussian process when the
coupling with the environment is weak or strong. In the first case 
decoherence is not much detrimental and for long interaction times the probe acquires
information about the environmental parameters without being too much
mixed. Conversely, for strong coupling information is quickly impinged on the quantum probe
and can effectively retrieved by measurements performed in the early
stage of the evolution.  In the intermediate situation, none of the two
above effects take place: information is flowing from the environment to
the probe too slowly compared to decoherence, and no measurements can be
effectively employed to extract it from the quantum probe.
The two regimes of weak and strong coupling are defined in terms of a
threshold value of the coupling, which itself increases with the fractional 
dimension. 
\end{abstract}
\end{frontmatter}
%
\section{Introduction}
Stochastic modelling is often the most effective tool available in order
to describe complex systems in physical, biological and social networks
\cite{Sib13,Wil09,Mos10,Foc10}.  In particular, since natural noise
sources are mostly Gaussian, stationary and non-stationary Gaussian
processes are often used to model the response of a system exposed to
environmental noise.  In view of the increasing interest towards complex
systems, a question thus naturally arises on whether an effective
characterization of Gaussian processes is achievable.
\par
In this paper we address the characterization of classical random fields
and focus attention on fractional Gaussian processes. The reason is
twofold: On the one hand, most of of the noise sources in nature are
Gaussian and the same is true for the linear response of systems exposed
to environmental noise \cite{Fox78}. 
On the other hand, fractional processes have
recently received large attention since they are suitable to describe
noise processes leading to complex trajectories, e.g. irregular time
series characterized by a Haussdorff fractal dimension in the range
$1\leq \delta \leq 2$.  In particular, in order to maintain the
discussion reasonably self contained, we focus on systems exposed to
fractional Brownian noise \cite{Man68,Man69,Taq13} (fBn) $B_H(t)$, which is a
paradigmatic nonstationary Gaussian stochastic process with zero mean
$E[B_H(t)]_B=0$ and covariance \cite{Bar88}
\begin{align}
E[B_H(t)B_H(s)]_B &\equiv K(t,s) \notag \\ 
&= \frac12 V_H\left(|t|^{2H} + 
|s|^{2 H} - |t-s|^{2H} \right)\,,
\label{BH}
\end{align}
where 
$$
V_H = \Gamma (1-2H)\, \frac{\cos \pi H}{\pi H}\,,
$$
$\Gamma(x)$ being the Euler Gamma function.
In the above formulas $H$ is a real parameter $H\in [0,1]$, usually
referred to as the Hurst parameter \cite{Hur51}.  The Hurst parameter is
directly linked to the fractal dimension $\delta=2-H$ of the
trajectories of the particles exposed to the fractional noise. 
The
notation $[...]_B$ denotes expectation values taken over the values of
the process and represents a shorthand for the functional integral
\begin{align}
[f(t)]_B&=\int\! {\cal D}[B_H(t)]\, {\cal P}[B_H(t)]\, f(t) 
\notag\\ &1 =\int\! {\cal D}[B_H(t)]\, {\cal P}[B_H(t)] \,,
\notag
\end{align}
performed over all the possible realizations of the process $B_H(t)$,
each one occurring with probability ${\cal P}[B_H(t)] $. We remind that
fBn is a self-similar Gaussian process, i.e.  $B_H(at) \sim |a|^H
B_H(t)$, and that it is suitable to describe anomalous diffusion
processes with diffusion coefficients proportional to $t^{2H}$,
corresponding to (generalized) noise spectra with a powerlaw
dependence $|\omega|^{-2H-1}$ on frequency \cite{Fla89}.
\par
The characterization of fBn amounts to the determination of the fractal
dimension of the resulting trajectories, i.e. the determination of the
parameter $H$. In the following, in order to simplify notation and formulas, we
will employ the complementary Hurst parameter $\gamma=1+H=3-\delta\in [1,2]$ and
upon replacing 
\begin{align}
H & \longrightarrow  \gamma-1 \notag \\
V_H & \longrightarrow V_\gamma=\frac2{\pi}\, 
\Gamma(2-2\gamma)\notag  \, \cos\pi\gamma 
\end{align}
in Eq. (\ref{BH}), we will denote the fBn process by $B_\gamma(t)$ 
\par
The purpose of this paper is to address in some details the
characterization of fBn, i.e. the determination of the parameter
$\gamma$, using {\em quantum probes}.  This means that we consider a
system, say a particle, subject to fBn, and assume that its motional degree
of freedom, regarded to be classical, is coupled to a quantum degree of
freedom of the same system, e.g. its spin. We then ignore the noisy
classical part and exploit quantum limited measurements on the spin part
to extract information about the fBn. 
Notice, however, that our approach and our results are also valid  to assess 
the performances of quantum limited measurements for any two-level system
subject to dephasing perturbations described by fractional Brownian 
noise, i,e. without the need of referring to a qubit coupled to the motion of a particle.
\par
We will address
both {\em estimation} and {\em discrimination} problems for the fractal dimension of the
fBn, i.e. situations where the goal is to estimate the unknown values
of the parameter $\gamma \in [1,2]$, and cases where we know in advance
that only two possible values $\gamma_1$ and $\gamma_2$ are admissible
and want to discriminate between them \cite{Dav87}.
\par
Several techniques have been suggested for the estimation of the Hurst 
parameter in the time or in the frequency domain \cite{Jeo07,Bar10}, 
or using wavelets \cite{Wor92,Zun07}.
Among them we mention 
range scale estimators \cite{Hur51}, maximum likelihood \cite{Ken99}, Karhunen-Loeve 
expansion \cite{Sal13},  p-variation \cite{Mag13}, periodograms
\cite{Taq95,Yin09}, weigthed functional \cite{Boy13}, and 
linear Bayesian models \cite{Mak11}.
\par
Compared to existing techniques, quantum probes offers the advantage of 
requiring measurements performed at a fixed single (optimized)
instant of time, without the need of observing the system for a long
time in order to collect a time series, and thus avoiding any issue related
to poor sampling \cite{Sch95,Meh97,Cas02}.
As we will see, quantum probes may be
effectively employed to characterize fractional Gaussian process when
the the system-environment coupling is weak, provided that a long
interaction time is achievable, or when the coupling is strong and the
quantum probe may be observed shortly after that the interaction has
been switched on. Overall, and together with results obtained for 
the characterization of stationary process \cite{ben14},
our results indicate that quantum probes
may represent a valid alternative to other techniques to characterize
classical noise.
\par
The paper is structured as follows: In Section \ref{s:mod} we introduce
the physical model and discuss the dynamics of the quantum probe. In
Section \ref{s:qest} we briefly review the basic notions of quantum
information geometry and evaluate the figures of merit that are 
relevant to our problems. In Section \ref{s:qp} we discuss optimization 
of the interaction time, and evaluate the ultimate bounds to the above 
figures of merit that are achievable using quantum probes. Section \ref{s:out}
closes the paper with some concluding remarks.
\section{The physical model} \label{s:mod}
We consider a spin $\frac12$ particle in a situation where its motion is
subject to environmental fBn noise and may be described classically. We
assume that the motional degree of freedom of the particle is coupled to
its spin, such that the effects of noise influence also the dynamics of the
spin part. We also assume that the noise spectrum of the fBn contain
frequencies that are far away from the natural frequency $\omega_0$ of
the spin part. When the spectrum contains frequencies that are {\em
smaller} than $\omega_0$ than the fluctuation induced by the fBn are 
likely to produce decoherence of the spin part, rather the damping, such that 
the time-dependent interaction Hamiltonian between the motional and the 
spin degrees of freedom may be written as 
\begin{align} 
\label{hi}
H_I = \lambda\, \sigma_z B_\gamma(t)\,,\end{align}
where $\sigma_z$ denotes a Pauli matrix and $\lambda$ denotes 
the coupling between the spin part and its classical environment. 
We do not refer to any specific interaction model between the motional degree of freedom and 
the spin part and assume that Eq. (\ref{hi}) describes the overall effect of the coupling.
The full Hamiltonian
of the spin part 
is given by $H=\omega_0 \sigma_z + \lambda B_\gamma(t) \sigma_z$ and may be easily
treated in the interaction picture. Upon denoting by $\rho_0$ the
initial state of the spin part, the state at a subsequent time $t$ is
given by 
$\rho_\gamma (t) = E\left[ U(t)\, \rho_0\, U^\dag (t) \right]_B$, 
where
\begin{align}\label{Udc}
U (t) &= \exp\left\{-i \lambda \int_0^t\! ds\, B_\gamma (s) \sigma_z \right\} \equiv 
e^{-i \varphi(t)\sigma_z} \\ & = \cos\varphi(t) {\mathbbm I} - i
\sin\varphi(t) \sigma_z\,.\notag
\end{align}
Upon substituting the above expression of $U(t)$ in $\rho_\gamma$ we arrive
at 
\begin{align}
\rho_\gamma (t) = &
E[\cos^2\varphi(t)]_B\, \rho_0 + 
E[\sin^2\varphi(t)]_B\, \sigma_z\rho_0\sigma_z
\notag \\ &- i E[\sin\varphi(t)\cos\varphi(t)]_B\,[\sigma_z,\rho_0] \notag \\
 = &p_\gamma (t,\lambda)\, \rho_0 + [1-p_\gamma(t,\lambda)]
\,\sigma_z\rho_0\sigma_z\,.\label{evol}
\end{align}
In writing the last equality, we have already employed the
averages over the realizations of the fractional process
\begin{align}
p_\gamma (t,\lambda) \equiv E[\cos^2\varphi(t)]_B &= \frac12
\left[1+\exp\left\{-\frac{\lambda\, t^{2\gamma} V_\gamma}{\gamma}\right\}\right]
\notag \\
E[\cos\varphi(t)\sin\varphi(t)]_B &= 0\notag\,,
\end{align}
which have been evaluated taking into account that $B_\gamma(t)$ is a Gaussian
process with zero mean and covariance $K(t,s)$, i.e. by using the
generating function
\begin{align}
E&\left[
  \exp\left\{-i \int_0^t\! ds\, f(s)\, B_\gamma(s)\right\} 
\right]_B =\notag\\   
& \exp\left\{-\frac12\int_0^t\!\int_0^t\! ds ds^\prime f(s)
\,K(s,s^\prime)\, f(s^\prime)\right\}\,,
\end{align}
which leads to
\begin{align}
E\left[ e^{-i m \varphi(t)}\right]_B & = E\left[
\exp\left\{-i m \int_0^t\!\! ds\, B_\gamma(s)\right\} 
\right]_B \notag \\ &= \exp\left\{-\frac12 m^2 \beta(t)
\right\} \quad \forall m\in {\mathbbm Z}\,, \notag 
\end{align}
where
\begin{align}
\beta(t) &= 
\int_0^t\!\int_0^t\! ds ds^\prime
\,K(s,s^\prime) = \frac{\lambda\, t^{2\gamma}}{2\gamma}V_\gamma\,.
\label{beta}
\end{align}
In the complementary case, i.e. when the noise spectrum of the fBn
contains frequencies that are larger than the natural frequency of the
spin part, the dominant process induced by the environmental noise is 
damping, such that the overall Hamiltonian may be written as 
$H^\prime=\omega_0 \sigma_z + B_\gamma (t) \sigma_x$. 
Due to the presence of the transverse field in the time-dependent stochastic
Hamiltonian there is no exact (close) solution for the unitary 
evolution, which involves time ordering. When the quantity $\beta(t)$ in the 
characteristic function is  small \cite{fullH}, e.g. in the limit of slowly varying 
$B_\gamma(t)$ we may write the quasi static unitary evolution, which 
reads as follows 
\begin{align}\label{Ud}
U^\prime(t) = &\exp\left\{-i \int_0^t \! ds\, H^\prime(s)\right\} =
\notag \\
= &\cos\sqrt{\omega_0^2 t^2 + \varphi^2(t)}\, {\mathbbm I}
- i \omega_0 t \frac{\sin\sqrt{\omega_0^2 t^2 + \varphi^2(t)}}{
\sqrt{\omega_0^2 t^2 + \varphi^2(t)}}\, 
\sigma_z \notag \\
&
- i \varphi(t) \frac{\sin\sqrt{\omega_0^2 t^2 + \varphi^2(t)}}{
\sqrt{\omega_0^2 t^2 + \varphi^2(t)}}\, 
\sigma_x \notag \\
 \simeq &\cos\varphi(t)\, {\mathbbm I} - i \sin\varphi(t)\,\sigma_x\,,
\end{align}
where the last equality is valid if $\omega_0 t\ll \varphi(t)$, i.e.
assuming $\omega_0\ll \lambda |B_\gamma(t)|$, $\forall t$. In this
limit, the damping evolution operator in Eq. (\ref{Ud}) is just a rotated
version of the decoherence one in Eq. (\ref{Udc}). In general
In the following we 
limit ourselves to estimation and discrimination problems involving a fBn
inducing nondissipative decoherence, i.e. with noise spectrum containing
frequencies smaller than $\omega_0$ and leading to an evolution operator
of the form (\ref{Udc}).
\section{Quantum information geometry for a spin $\frac12$ particle
exposed to classical noise} \label{s:qest}
The characterization of fBn by quantum probes amount to distinguish 
quantum states in the class $\rho_\gamma(t)$, i.e. states originating 
from a common initial state $\rho_0$ and evolving in different noisy 
fBn channels, each one characterized by a different Hurst parameter, and
thus inducing trajectorie with different fractal dimension.
Distinguishability of quantum states is generally quantified by a distance in
the Hilbert space. However, depending on the nature of the
estimation/discrimination problem at hand, different distances are
involved to capture the relevant notion of distinguishability
\cite{LNP649,gqs}.
\par
In situations where we want to estimate the unknown value of $\gamma \in
[1,2]$ the problem is to discriminate a quantum state within the
continuous family $\rho_\gamma(t)$. In this case, the relevant quantity
is the so-called Bures infinitesimal distance between nearby point in
the parameter space 
\cite{bur69,uhl76,woo81,jos94,som03}
$d^2_B(\rho_\gamma,\rho_{\gamma+d\gamma}) = g_B (\gamma)\, 
d^2\gamma$, 
where the {\em Bures metric} $g_B(\gamma)$ is given by 
\begin{align}\label{gb}
g_B(\gamma) = \frac12 \sum_{nk}
\frac{|\langle\psi_k|\partial_\gamma\rho_\gamma|\psi_k\rangle|^2}{\rho_n+\rho_k}\,,
\end{align}
$|\psi_n\rangle$ being the eigenvectors of $\rho_\gamma=\sum_n \rho_n
|\psi_n\rangle\langle\psi_n |$. We omitted the explicit dependence on time.
The finite Bures distance between two quantum states
is given by $D_B(\rho_1,\rho_2)^2=2 (1-\sqrt{F(\rho_1,\rho_2)})$ in
terms of the fidelity $F(\rho_1,\rho_2) =
\left(\hbox{Tr}\left[\sqrt{\sqrt{\rho_1}\rho_2\sqrt{\rho_1}}\right]\right)^2$.
\par 
The relevance of the Bures metric in estimation problems comes from the
fact that $g_B(\gamma) = \frac14 G(\gamma)$ where $G(\gamma)$ is the
quantum Fisher information of the considered statistical model
$\rho_\gamma$ \cite{BC94,BC96,bro9X,nag00,Zan08,Inv08,Par09}. 
In order to appreciate this fact, let us remind that 
any estimation problem consists in inferring
the value of a parameter $\gamma$, which is not directly accessible, 
by measuring a related quantity $X$. 
The solution of the problem amounts to find an estimator $\hat{\gamma}
\equiv \hat\gamma
(x_1, x_2, \ldots)$, {\em i.e.} a real function of the measurements
outcomes $\{x_k\}$ to the parameters space.  Classically, the variance
$\hbox{Var}(\gamma)$ of any 
unbiased estimator satisfies the Cramer-Rao bound 
$\hbox{Var}(\gamma)\geq 1/{M F(\gamma)}$, 
which establishes a lower bound on variance in terms of the number of independent
measurements $M$ and the Fisher Information $F(\gamma) =\sum_x p(x |
\gamma) \left [\partial_\gamma \log p(x | \gamma)\right ]^2$,
$p(x | \gamma)$ being the conditional probability of
obtaining the value $x$ when the parameter has the value $\gamma$.
When quantum systems are involved, we have 
$p(x | \gamma) =\hbox{Tr}\left[\varrho_\gamma\,P_x\right]$, $\{P_x\}$
being the probability operator-valued measure (POVM) describing the
measurement. A quantum estimation problem thus corresponds to 
a quantum statistical model, {\em i.e.} a set of quantum states 
$\rho_\gamma$ labeled by the parameter of interest, with the mapping
$\gamma\to\rho_\gamma$ providing a coordinate system.
Upon introducing the Symmetric Logarithmic Derivative (SLD) $\Lambda_\gamma$
as operator satisfying the equation 
$\partial_\gamma\rho _\gamma=\frac{1}{2} \Big [ \Lambda_\gamma \rho _\gamma 
+ \rho _\gamma\Lambda_\gamma \Big ]$ 
one can prove \cite{BC94} that $F (\gamma)$ is upper bounded by the 
Quantum Fisher Information 
$F(\gamma) \leq G(\gamma)\equiv \hbox{Tr} \left [ \rho_\gamma
\Lambda_\gamma^2\right ]$.
In turn, the ultimate limit to precision is given by the quantum
Cramer-Rao theorem (QCR) 
$$\hbox{Var}(\gamma)\geq \frac{1}{M G(\gamma)}\:,$$
which provides a measurement-independent lower bound for the variance 
which is attainable upon measuring a POVM built with the
eigenprojectors of the SLD.
In fact, quantum estimation theory has been successfully 
employed for the estimation of static noise parameters 
\cite{hotta05,monras07,fuj01,ji08,dau06} and in several other
scenarios, as for example quantum thermometry \cite{brunelli2011}. 
\par
For quantum systems with a bidimensional Hilbert space, as those we are
investigating in this paper, the optimal measurement is a projective one
\cite{gil00,lua04}. Besides, using Eqs. (\ref{evol}) and (\ref{gb}), 
it is straightforward to show that  starting from a generic pure initial
state $|\psi_0\rangle=\cos\frac{\theta}{2} |0\rangle + e^{i\phi}
\sin\frac{\theta}{2} |1\rangle$ the maximum of $g_B(\gamma)$ is
achieved for $\theta=\pi/2$. In this case, the evolved state 
$\rho_\gamma(t)$ is a mixed state with 
eigenvectors independent on $\gamma$. In other
words, the dependence on $\gamma$ is only in the eigenvalues, and thus 
Eq. (\ref{gb}) reduces to
\begin{align}
g_B(\gamma) =& \frac14 \frac{\left[\partial_\gamma
p_\gamma(t,\lambda)\right]^2}{p_\gamma(t,\lambda)[1-p_\gamma(t,\lambda)]} 
\notag \\ =&
\frac{t^{4 \gamma}\,\lambda^2}{\gamma^4}
\left[\gamma\, \partial_\gamma V_\gamma -
 (1-2\gamma \log t)\,V_\gamma
\right]^2 \notag \\ 
& \times
\left(e^{\frac{2\lambda\,t^{2\gamma}}{\gamma}V_\gamma}-1\right)^{-1}\,,
\label{gbg}
\end{align}
where 
$$
\partial_\gamma V_\gamma= - \frac2{\pi} \Gamma(2-2\gamma) \Big[ 2 \cos
\pi\gamma\, \psi(2-2\gamma) + \pi \sin \pi \gamma\big]\,,$$
$\psi(x)=\partial_x \Gamma(x)/\Gamma(x)$ being the 
the log-derivative of the Euler Gamma function. 
\par 
The quantum Cramer-Rao theorem implies that the optimal conditions to
estimate $\gamma$ by quantum probes correspond to the maxima of
$g_B(\gamma)$. As mentioned above, the optimization over the initial state is trivial
and correspond to prepare the spin of the particle in the superposition
$|\psi_0\rangle=(|0\rangle+|1\rangle)/\sqrt{2}$, whereas the
maximization over the time evolution will be discussed in the next
Section.
\par
Let us now consider situations where we have to discriminate between 
two fixed and known values of $\gamma$, e.g. the null hypothesis 
$\gamma_1=2$ and the alternative $\gamma_2=\gamma^*$ corresponding 
to a non trivial fractal dimension. The corresponding states
$\rho_{\gamma_1}$  and $\rho_{\gamma_2}$ are assumed to be known, as well as the 
{\em a priori} probabilities $z_1$ and $z_2=1-z_1$, but we don't know which state is 
actually received at the end of propagation.
The simplest case occurs when the {\em a priori} probabilities are 
equal $z_1=z_2=\frac12$.
Any strategy for the discrimination between the two states 
amounts to define a two-outcomes POVM $\{\Pi_1,\Pi_2\}$ on the system
and establish the inference rule that after observing the 
outcome $j$ the observer infers that the state of the system is
$\rho_{\gamma_j}$ \cite{Helstrom,rv0,rv1,rv2,rv3}. 
The probability of inferring  $\gamma_j$ when the true value is $\gamma_k$ 
is thus given by $P_{jk} = \hbox{Tr}\left[\rho_{\gamma_k} \Pi_j\right]$ and the optimal 
POVM for the discrimination problem is the one minimizing the overall 
probability of a misidentification i.e. $P_e = z_1 P_{21} + z_2 P_{12}$. 
For the simplest case of equiprobable hypotheses ($z_1=z_2=1/2$) we have
$P_e = \frac{1}{2} \left(1-\Tr \left[\Pi_2 \Lambda\right]\right)$ where 
$\Lambda=\frac12(\rho_2-\rho_1)$. $P_e$ is minimized by 
choosing $\Pi_2$ as the projector over the positive subspace of $\Lambda$. 
Then we have $\Tr [\Pi_2\Lambda] =\Tr |\Lambda| $ and 
$P_e = \frac{1}{2}\left ( 1- \Tr \left|\Lambda \right|\right)$
where $|A|=\sqrt{A^\dag A}$. This is usually referred to as the
Helstrom bound, and represent the ultimate quantum bound to the error
probability in a binary discrimination problem.
In our case, $P_e$ is minimized when the two
output states commute, i.e. for $\theta=\pi/2$ leading to
\begin{align}
P_e 
&= \frac12 \left(1-\left|p_{\gamma_2}(t,\lambda) - 
p_{\gamma_1} (t,\lambda)\right|\right) \notag \\ 
&= \frac12 \left(1-\frac12\left|e^{-2\beta_1 (t)}- 
e^{-2\beta_2(t)}\right|\right) \label{peqp}
\end{align}
where $p_\gamma (t,\lambda) = \frac12 (1+e^{-2\beta(t)})$ is
given in Section \ref{s:mod}.
The minimization over the interaction time will be discussed 
in the next Section. We notice, however, that any single-copy
discrimination strategy based on quantum probes is inherently 
inefficient since Eq. (\ref{peqp}) imposes an error probability 
larger than $P_e\geq \frac14$ at any time. One is therefore led 
to consider different strategies, as those involving several copies
of the quantum probes.
\par
Indeed, let us now suppose that $n$ copies of both states are available for 
the discrimination. The problem may be addressed
using the above formulas upon replacing $\rho$ with $\rho^{\otimes n}$. 
We thus need to analyze the  quantity $P_{e,n} = \frac{1}{2}\left ( 1
-  \Tr \frac12(|\rho_{\gamma_2}^{\otimes n} -\rho_{\gamma_1}^{\otimes
   n})|\right )$. 
The evaluation of the trace distance for increasing $n$ may be 
difficult and for this reason, one usually resort to the quantum
Chernoff bound, which gives an upper bound to the
probability of error \cite{PRAQCB,PRLQCB,Nuss,Aud,Pirandola,Pir12}
$$P_{e,n} \leq \frac 12 Q^n$$
where 
\begin{align}\label{Q}
Q\equiv Q[\gamma_1,\gamma_2,\lambda]= 
\inf_{0\leq s\leq 1} \Tr\left[\rho_{\gamma_1} ^s\:
\rho_{\gamma_2}^{1-s}\right]\,.\end{align}
The bound may be attained in the asymptotic limit of large $n$.
Notice that while the trace distance is capturing the notion of 
distinguishability for single copy discrimination this is not 
the case for multiple copies strategies, where the quantity $Q$ 
represent the proper figure of merit. 
Also in Eq. (\ref{Q}) we omitted the explicit dependence on the
interaction time.
\par
For nearby states the relevant distance is the so-called
infinitesimal quantum Chernoff bound (QCB) distance 
$d^2_{QCB}(\rho_\gamma,\rho_{\gamma+d\gamma}) =
1 - Q
= g_{QCB} (\gamma)\, 
d^2 \gamma$, where the QCB metric $g_{QCB}(\gamma)$ is given by
\begin{align}\label{gqcb}
g_{QCB}(\gamma) = \frac12 \sum_{nk}
\frac{|\langle\psi_k|\partial_\gamma\rho_\gamma|\psi_k\rangle|^2}
{\left(\sqrt{\rho_n}+\sqrt{\rho_k}\right)^2}\,.
\end{align}
The QCB introduces a measure of distinguishability for density operators
which acquires an operational meaning in the asymptotic limit. The
larger is the QCB distance, the smaller is the asymptotic error
probability of discriminating a given state from its neighbors. On the
other hand, for a fixed probability of error $P_e$, the smaller is $Q$, 
the smaller the number of copies of $\rho_{\gamma_1}$ and
$\rho_{\gamma_2}$ we will need in order to distinguish them. 
\par
Also the quantity $Q$ is minimized when the two
output states commute, i.e. for $\theta=\pi/2$ and, in this case we have
\begin{align}\label{xiqcb}
Q= \inf_s
 & \Big\{p^s_{\gamma_1}(t,\lambda)
\,p^{1-s}_{\gamma_2}(t,\lambda)
 \\  &+[1-p_{\gamma_1}(t,\lambda)]^s
\,[1-p_{\gamma_2}(t,\lambda)]^{1-s}
\Big\}\,. \notag
\end{align}
The minimization over the parameter $s$ and the interaction time will be discussed 
in the next Section.
Concerning the QCB metric, we have the general relation $\frac12
g_B(\gamma) \leq g_{QCB}(\gamma)\leq g_B(\gamma)$. In our case, since
the maximum is achieved when only the eigenvalues of $\rho_\gamma (t)$ 
depends on $\gamma$, the only non zero terms in Eqs. (\ref{gb}) and
(\ref{gqcb}) are those with $n=m$. As a consequence the first inequality
above is saturated and we have $g_{QCB} (\lambda) = \frac12
g_B(\lambda)$, $\forall t, \gamma$. The working conditions to optimize
the estimation or the discrimination of nearby states are thus the same.
\section{Quantum probes for fractional Gaussian processes}\label{s:qp}
In this Section we discuss optimization of the estimation/discrimination
strategies for fBn over the possible values of the interaction time. More 
explicitly, we maximize the Bures metric and minimize the Helstrom and
QCB bound to error probability, as a function of the interaction time. In this way, 
we individuate the optimal working conditions, maximizing the performances 
of quantum probes, and establish a benchmark to assess any strategy
based on non optimal measurements.
\begin{figure}[h!]
\includegraphics[width=0.99\columnwidth]{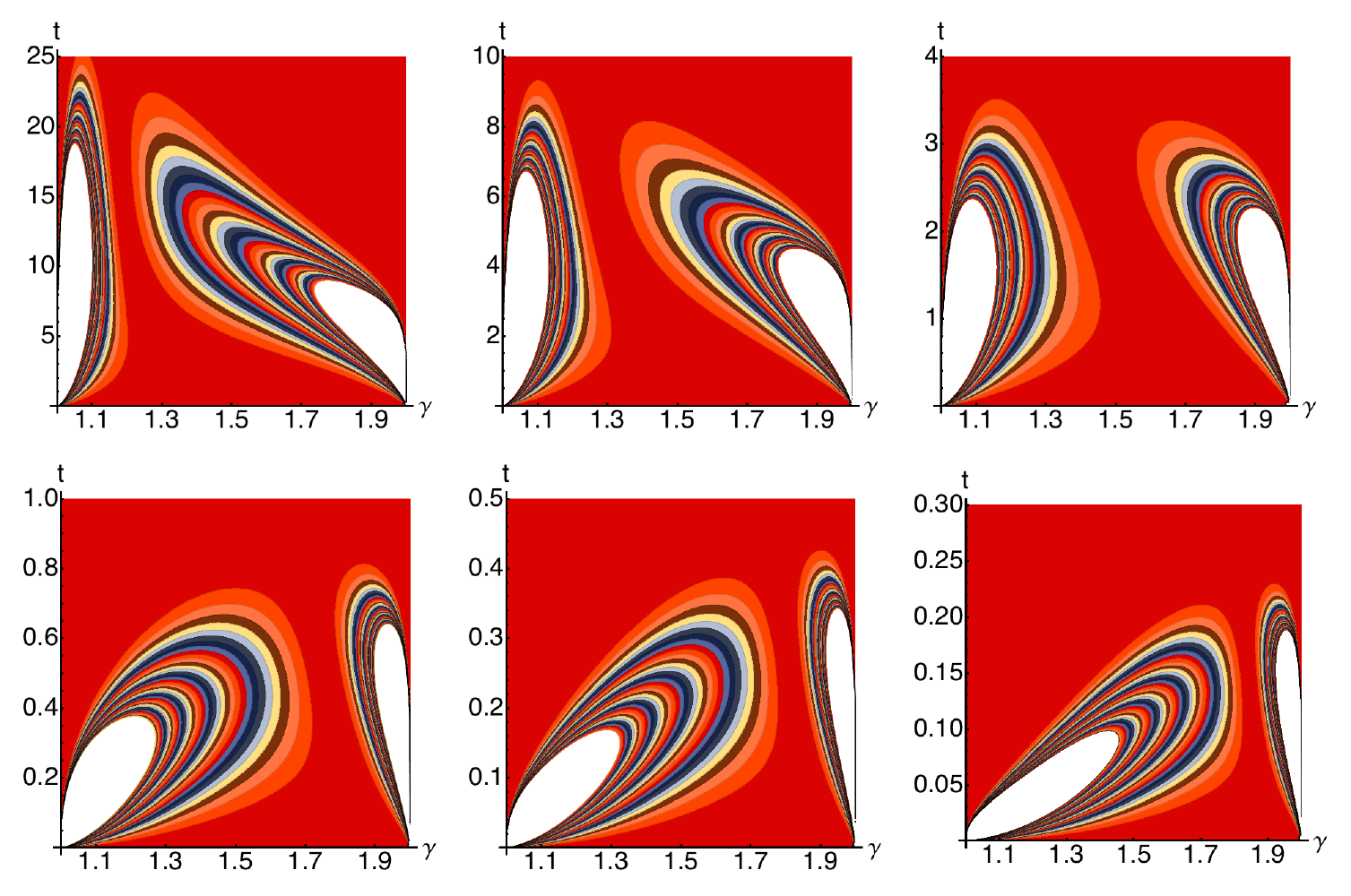}
\caption{Bures metric $g_B(\gamma)$ for the estimation of the complementary 
Hurst parameter $\gamma$ as a function of $\gamma$
and of the interaction time, for different values of the coupling
$\lambda$. The contour plots correspond, from top left to bottom right,
to $\lambda=10^{-3}, 10^{-2}, 10^{-1}, 10,
\label{f:f1}
10^2$, and $\lambda=10^3$, respectively. Whiter regions correspond to
larger values of the Bures metric.}\end{figure}
\par
\subsection{Estimation by quantum probes}
Upon inspecting the functional dependence of the Bures metric 
on the quantities $t$, $\lambda$ and $\gamma$ in Eq. (\ref{gbg})
one sees that $g_B(\gamma)$ is somehow a function of the 
quantity $\lambda t^{2\gamma}$ and thus maxima are expected, 
loosely speaking, for small $t$ and large $\lambda$ or viceversa.
On the other hand, this scaling is not exact and thus a richer 
structure is expected.
This is illustrated in Fig. \ref{f:f1}, where we show contour plots of 
$g_B$ as a function of $\gamma$ and of the interaction time for different 
values of the coupling $\lambda$. 
As it is apparent from the plots, for 
any value of the coupling there are two 
maxima located in different regions (notice the different ranges
for the interaction time). The global maximum moves from one region to
the other depending on the values of the coupling (see below). 
\par
In Fig. \ref{f:f2} we show the results obtained from the numerical 
maximization of the Bures metric $g_(\gamma)$ over the interaction 
time. The upper left panel is a log-log-plot of the maximized Bures 
metric as a function of the coupling for randomly chosen values of $\gamma\in[1,2]$ 
and $\lambda\in [10^{-3}, 10^3]$ (gray points). 
We also report some curves at fixed values of $\gamma$, showing
that for any value of the complementary Hurst parameter, except those
close to the limiting values $\gamma=1$ and $\gamma=2$, a threshold 
value $\lambda_{\rm th}(\gamma)$ on the coupling, i.e. on the intensity 
of the noise, naturally emerges. The Bures metric is large, i.e.
estimation may achieve high precision, in the weak and in the strong 
coupling limit, that is, when $\lambda\ll \lambda_{\rm th}(\gamma)$
or $\lambda\gg \lambda_{\rm th}(\gamma)$. On the other hand, for
intermediate values of the coupling 
$\lambda \sim \lambda_{\rm th}(\gamma)$ the estimation of the fractal
dimension is inherently inefficient. 
This behavior is further illustrated in the lower left panel, where
we report the same random points as a function of $\gamma$, also
showing curves at fixed values of the coupling. Values of $\gamma$ close
to $\gamma=1$ or $\gamma=2$ may be precisely estimated for any value of
the coupling whereas intermediate values needs a tuning of $\lambda$, in
order to be placed in the corresponding weak (or strong) coupling limit.
The threshold value $\lambda_{\rm th} (\gamma)$ increases with $\gamma$ 
and does not appear for $\gamma\simeq 1$ or $\gamma\simeq 2$. For those
values high precision measurements are achievable only in the strong 
coupling limit (for $\gamma\simeq 1$, i.e. fractal dimension close 
to $\delta\simeq2$) or the weak coupling limit 
($\gamma\simeq 2$, i.e. negligible fractal dimension $\delta\simeq 1$).
\par
\begin{figure}[h!]
\begin{tabular}{cc}
\includegraphics[width=0.47\columnwidth]{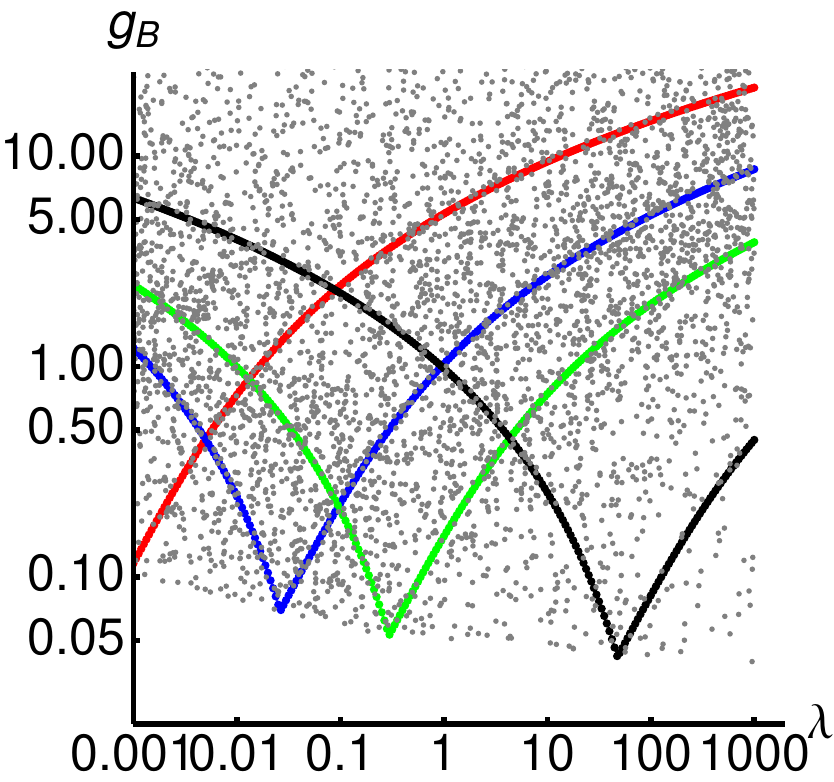}&
\includegraphics[width=0.47\columnwidth]{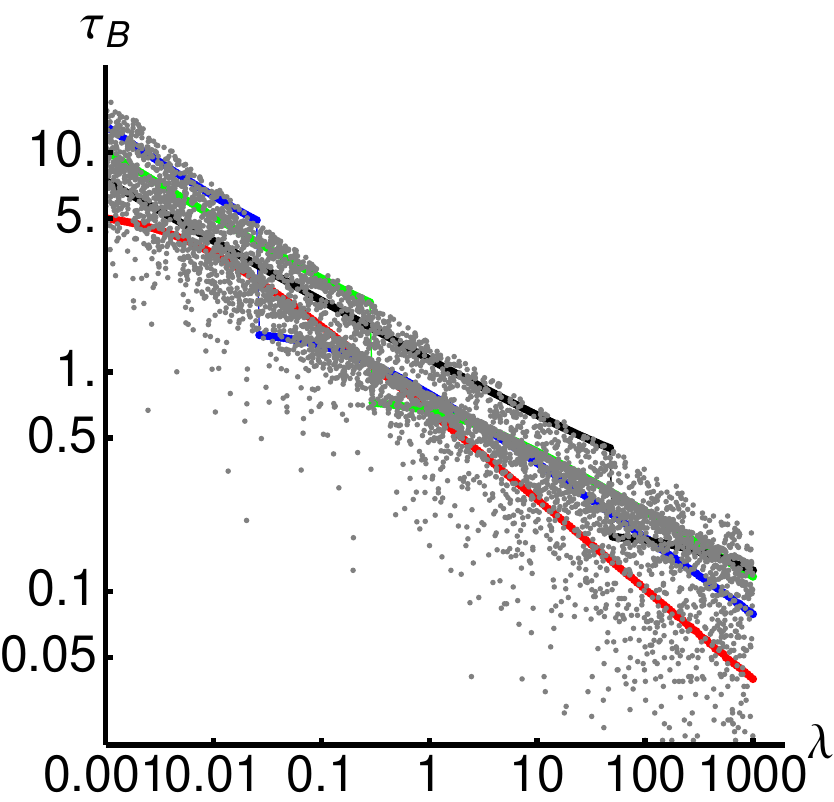}\\
\includegraphics[width=0.47\columnwidth]{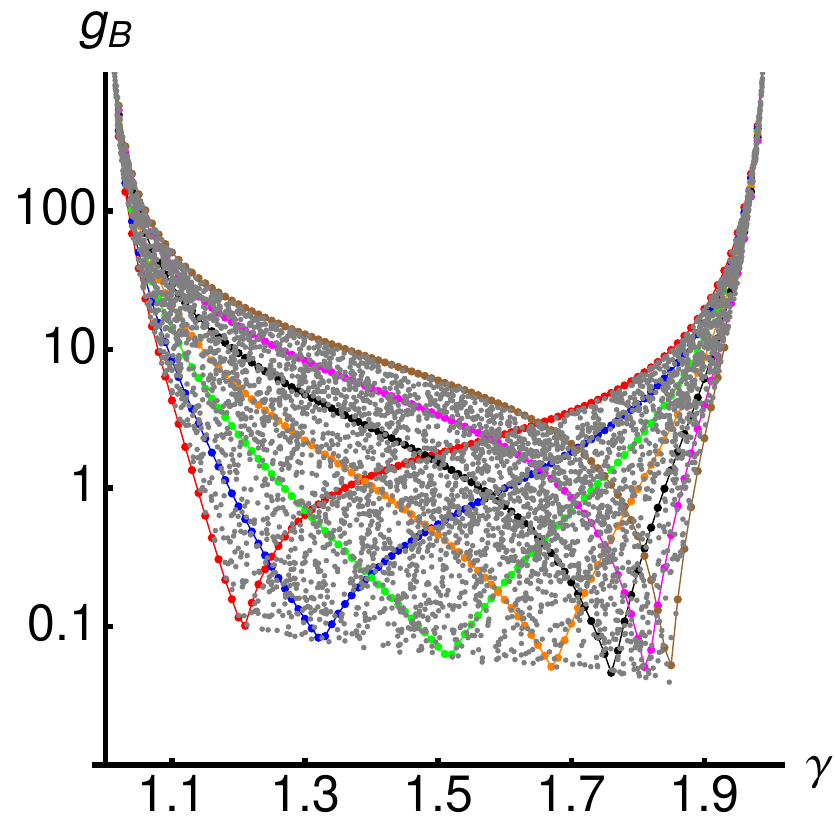}&
\includegraphics[width=0.47\columnwidth]{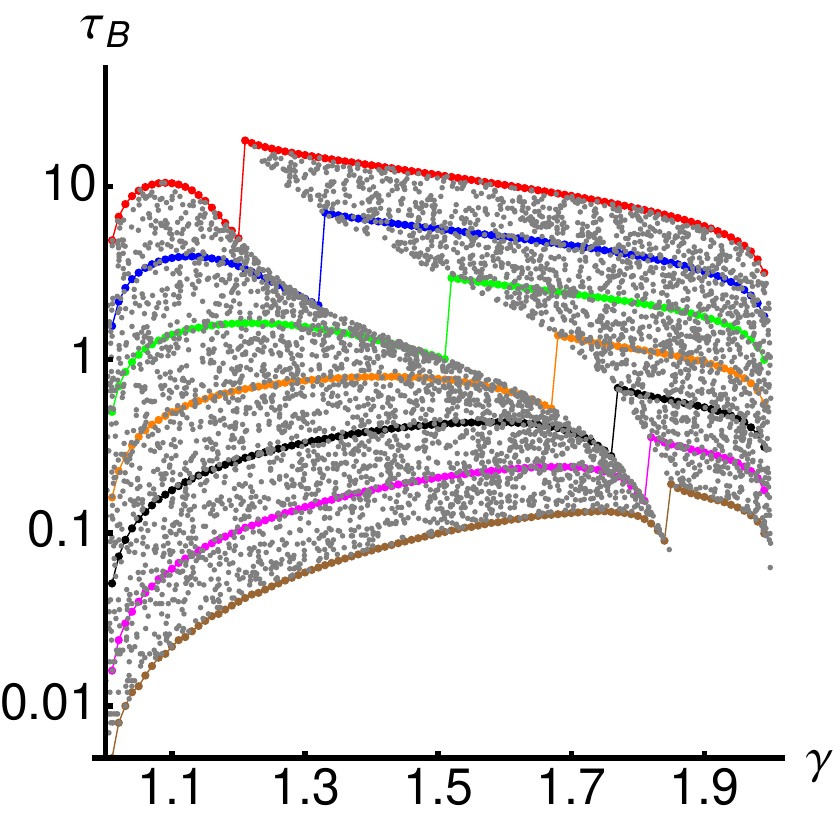}
\end{tabular}
\caption{Optimal estimation of the complementary Hurst parameter 
$\gamma$ by quantum probes. The upper left panel shows the maximized 
Bures metric as a function of the coupling for $5000$ randomly 
chosen values of $\gamma\in[1,2]$ (gray points) and $\lambda 
\in [10^{-3},10^3]$. The curves correspond to, from left to right, to 
$\gamma=1.2$ (red), $\gamma=1.4$ (blue) 
$\gamma=1.6$ (green), $\gamma=1.8$ (black).
The upper right panel shows the optimal values 
$\tau_B$ of the interaction time, leading to the 
Bures metric of the left panel. The curves are for the same 
fixed values of $\gamma$ of the left panel.
The lower left panel shows the the maximized 
Bures metric as a function of $\gamma$ for the same $5000$ 
randomly  chosen values of $\gamma\in[1,2]$ (gray points) 
and $\lambda \in [10^{-3},10^3]$ of the upper panel. Here we report
curves at fixed values of $\lambda=10^k$ with (from left to right)
$k=-3,-2,-1,0,1,2,3$. The lower right panel shows the optimal values of
the interaction time, leading to the Bures metric of the right panel.
\label{f:f2}
}
\end{figure}
\par
The right panels of Fig. \ref{f:f2} show the optimal values 
$\tau_B = \arg \max_t g_B (\gamma)$ of the interaction 
time, leading to the 
maximized values of the Bures metric reported in 
the left panels. The upper panel shows $\tau_B$ as function of the 
coupling whereas the lower one illustrates the behavior as a function 
of $\gamma$. Referring to the upper panel: $\tau_B$ exhibits a power-law
decrease for small and large values
of the coupling (notice the log-log scale the plots)
whereas for intermediate values of $\lambda$ we observe
a discontinuous behavior, which reflects the transition
of the global maximum from from the peak at large $t$ and small $\lambda$ to
the other one, located in the region of small $t$ and large
$\lambda$. 
\par
The overall picture that we obtain from Fig. \ref{f:f2} 
is that quantum probes may be generally employed to obtain 
a reliable characterization of fractional Gaussian process, 
except when the coupling with the environment has intermediate values.
These results may be understood intuitively as follows. 
The maxima obtained for small values of $\lambda$ correspond 
to quantum probes that are weakly coupled to the environment. 
In this case, decoherence is not much detrimental and for
long interaction times the probe acquires information
about the environmental parameters without being 
too much mixed, i.e. still storing this information in 
its quantum  state. Viceversa, for a quantum probe strongly 
coupled to the environment, the information about the environmental
parameters is quickly impinged onto the state of the quantum probe,
such that it can effectively retrieved, upon performing measurements 
in the early stage of the evolution.
In the intermediate situation, none of the two above effects
take place: information is flowing from the environment to the
probe too slowly compared to decoherence and no measurements
can be effectively employed to extract it from the quantum state
of the probe.
The two regimes of weak and strong coupling are defined in terms of a
threshold value of the coupling, which itself increases with the fractional 
dimension. 
\par
The above picture, however, does not apply when the fractal dimension
of the trajectories is close to its limiting values, i.e. when the 
complementary Hurst parameter assumes values close to $\gamma=1$
or $\gamma=2$. In these two limiting cases no threshold on the 
coupling appears and $\gamma$ may be reliably estimated only in the weak 
coupling limit (for negligible fractal dimension) or in the strong
coupling one (fractal dimension closer to its maximum value).
\subsection{Discrimination by quantum probes}
Let us now consider  discrimination
problems involving the complementary Hurst parameter. We assume
to know in advance that only two possible values 
$\gamma_1$ and $\gamma_2$ are admissible and want to discriminate
between them using the results of a measurement performed on
the quantum probe. The Helstrom bound $P_e$ to the error probability
in a single-shot discrimination is given in Eq. (\ref{peqp}) 
and here we want to minimize $P_e$ over the interaction time.
Results of the numerical minimization are shown in Fig. \ref{f:f2}, where
we report the minimized Helstrom bound as a function of $\gamma_2$
for different fixed values of $\gamma_1$, together with density plots 
of the same quantity as a function of the
pair of values $(\gamma_1, \gamma_2)$ for different values
of the coupling with the environment. 
\par
\begin{figure}[h!]
\begin{tabular}{ccc}
\includegraphics[width=0.31\columnwidth]{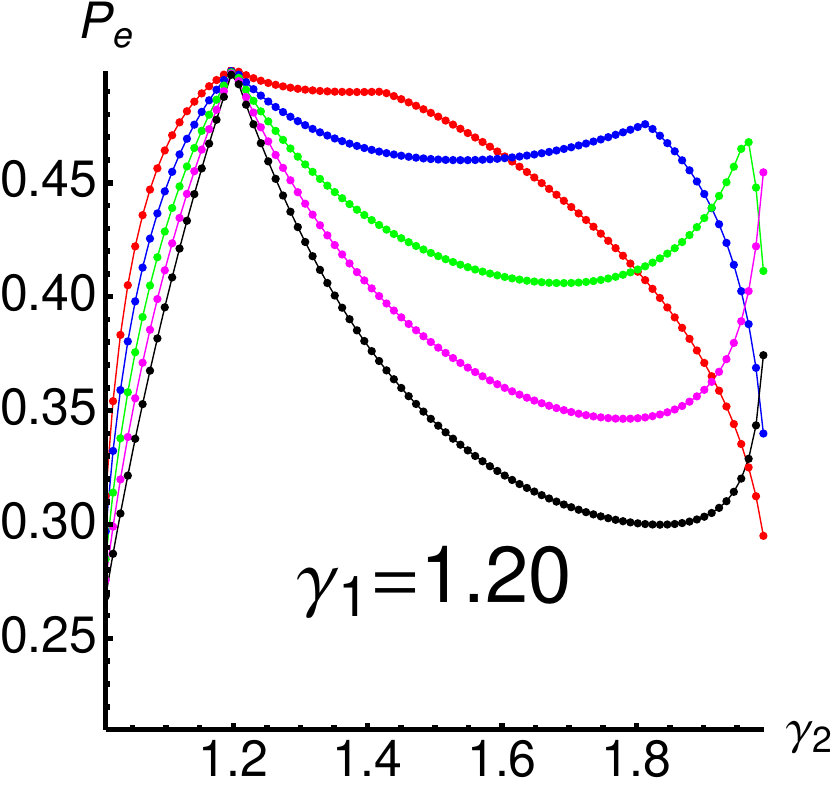}&
\includegraphics[width=0.31\columnwidth]{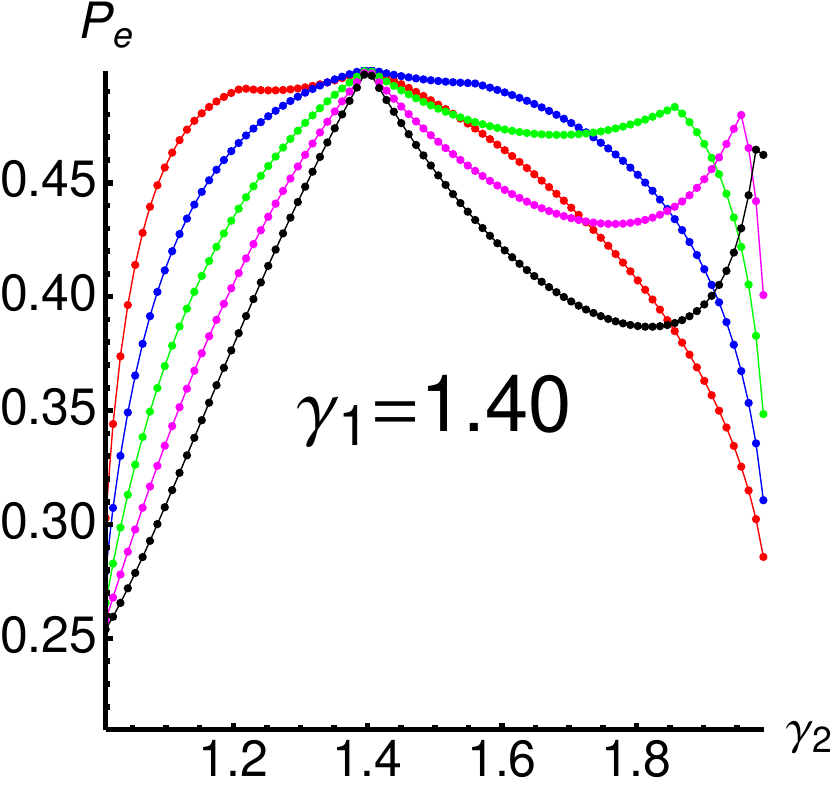}&
\includegraphics[width=0.31\columnwidth]{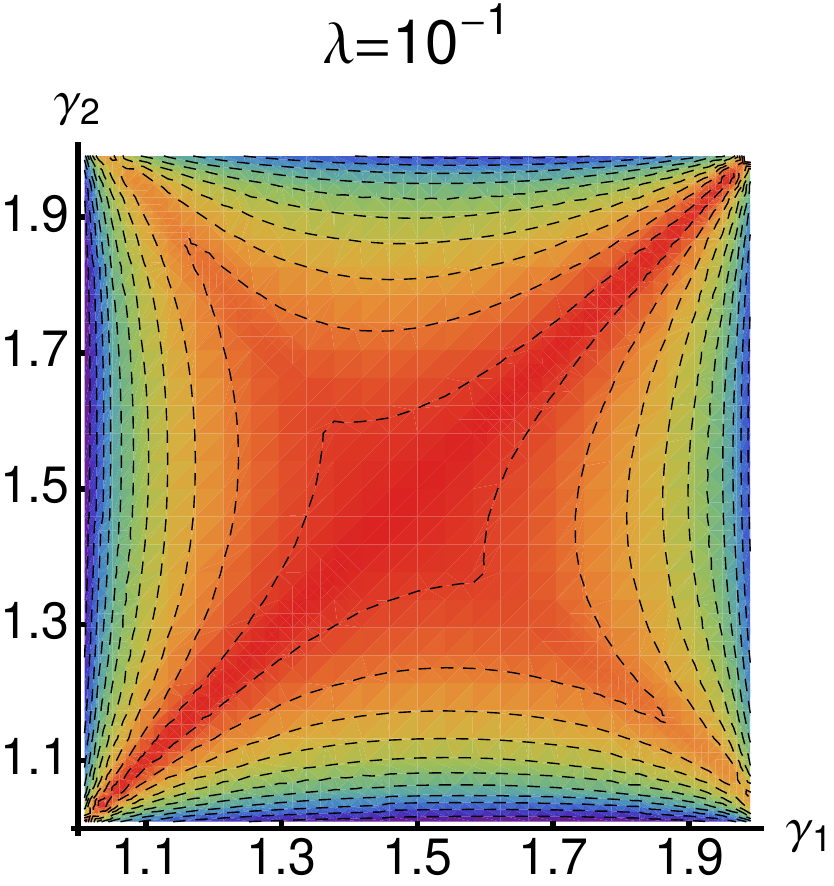}\\
\includegraphics[width=0.31\columnwidth]{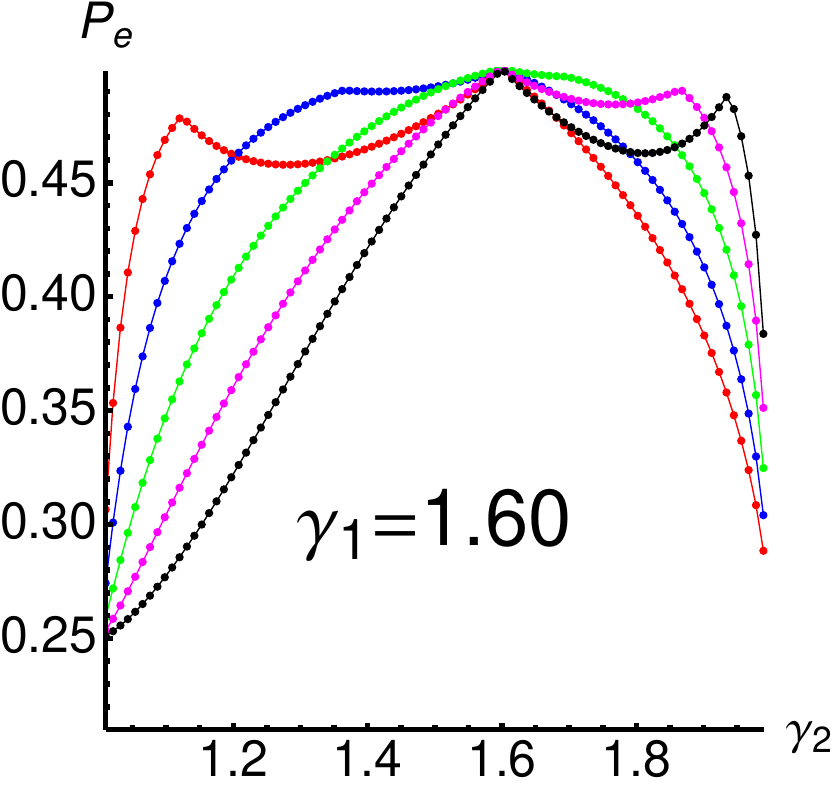}&
\includegraphics[width=0.31\columnwidth]{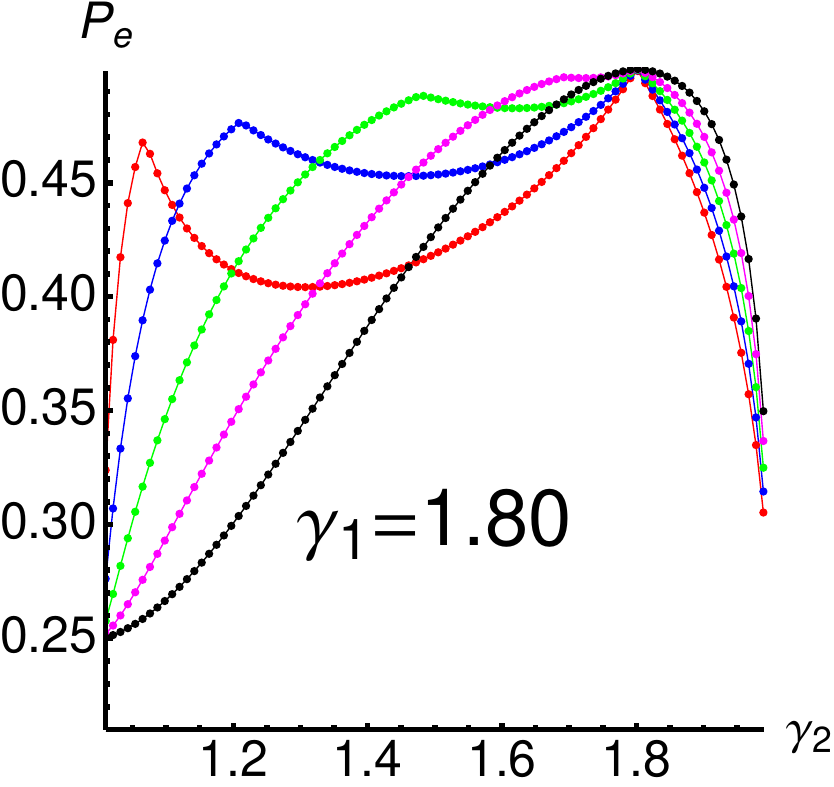}&
\includegraphics[width=0.31\columnwidth]{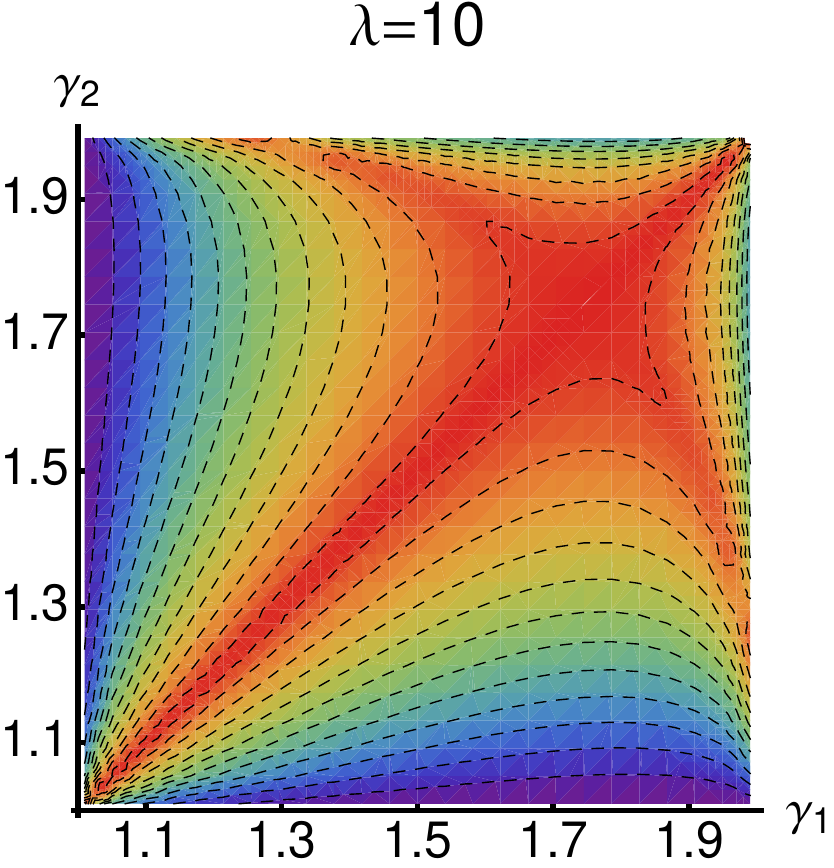}
\end{tabular}
\caption{Helstrom bound to the discrimination of pairs 
of values of the complementary Hurst parameter by quantum probes.
The four plots on the left panels show the Helstrom bound $P_e$ minimized over 
the interaction time as a function of $\gamma_2$ for different values
of $\gamma_1$. In all the plots the different curves refer to different
values of the coupling: $\lambda=10^{-2}$ (red), $\lambda=10^{-1}$
(blue), $\lambda=1$ (green), $\lambda=10$ (magenta), 
$\lambda=100$ (black). The two right panels show a density plot of the minimized
Helstrom bound as a function of both the values $\gamma_1$
and $\gamma_2$ for two different values of the coupling: 
$\lambda=10^{-1}$ (top panel) and $\lambda=10$ (bottom panel). Blue
regions correspond to smaller values of $P_e$.
\label{f:f3}}
\end{figure}
\par
The plots confirm the overall symmetry of the Helstrom bound 
$P_e(\gamma_1,\gamma_2)=P_e(\gamma_2,\gamma_1)$ at fixed $\lambda$.
Another feature that emerges from Fig. \ref{f:f3} is that, say, 
the pairs $\gamma_1=1.2$ and $\gamma_2=1.4$ or $\gamma_1=1.4$ 
and $\gamma_2=1.6$ have different 
discriminability despite the fact that for both pairs 
we have $|\gamma_1-\gamma_2|=0.2$, i.e. the Helstrom bound 
is not uniform. The plots also confirm the overall picture 
obtained in discussing estimation problems: for each pair
of values $(\gamma_1,\gamma_2)$, two 
regimes of strong or weak coupling
may be individuated, where discrimination may be performed
with reduced error probability, whereas for intermediate values of the coupling 
performances are degraded. The only exception regards
values close to the limiting values $\gamma=1$ or $\gamma=2$,
where no threshold appears.
\par
We also notice that by increasing the coupling one enlarges 
the region in the $\gamma_1$-$\gamma_2$ plane where discrimination
may be performed with reduce error probability.
This is illustrated in the right panels of Fig. \ref{f:f3}, where we show 
a density plot of the minimized
Helstrom bound as a function of both the values $\gamma_1$
and $\gamma_2$ for two different values of the coupling: 
$\lambda=10^{-1}$ (top panel) and $\lambda=10$ (bottom panel)
\par
\begin{figure}[h!]
\includegraphics[width=0.48\columnwidth]{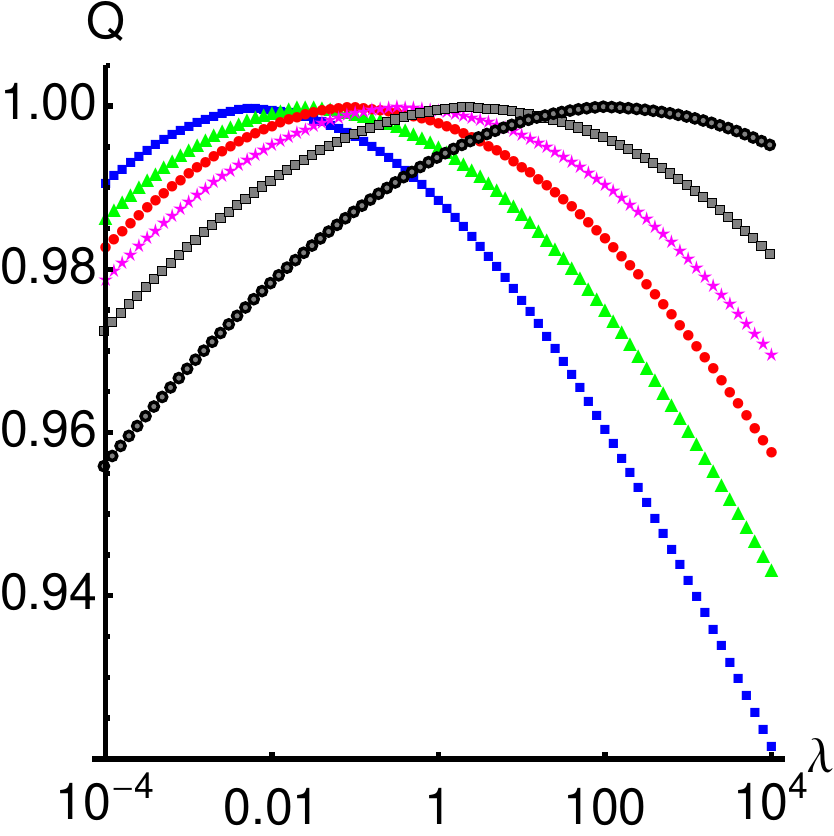}
\includegraphics[width=0.48\columnwidth]{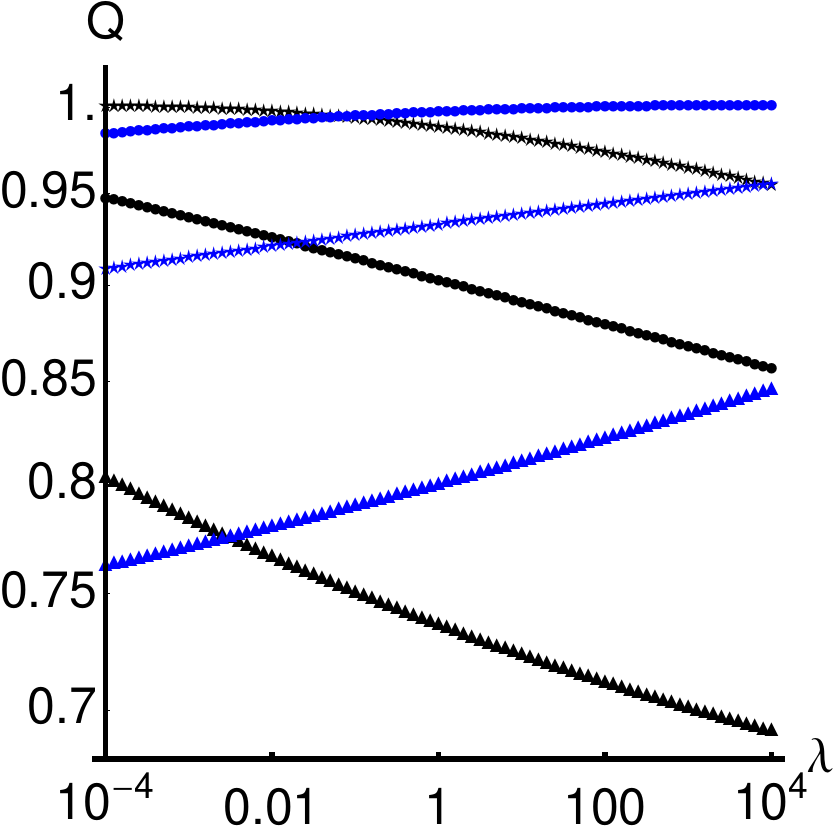}
\caption{Chernoff bound to the multiple-copy discrimination of pairs 
\label{f:f4}
of values of the complementary Hurst parameter by quantum probes.
In the left panel we report the maximized Chernoff bound as a function of the
coupling with the environment for pair of values $(\gamma,\gamma+0.2)$ 
with $\gamma$ not too close to the limiting values $\gamma=1$ or $\gamma=2$. 
From left to right we have, $\gamma=1.2$ (blue squares), $\gamma=1.3$ (green triangles), 
$\gamma=1.4$ (red circles), $\gamma=1.5$ (magenta stars), $\gamma=1.6$ (gray squares), 
$\gamma=1.7$ (gray circles).  In the right panel we show the same quantity for 
pair of values $(\gamma_1, \gamma_2)$
close to the boundaries $\gamma=1$ and $\gamma=2$. The increasing curves
correspond to $\gamma_1=1.0, \gamma_2=1.1$ (blue circles), 
$\gamma_1=1.1, \gamma_2=1.2$ (blue stars), $\gamma_1=1.0, \gamma_2=1.2$ (blue triangles), 
whereas the decreasing ones are for $\gamma_1=1.8, \gamma_2=1.9$ (black circles), 
$\gamma_1=1.9, \gamma_2=2.0$ (black stars), $\gamma_1=1.8, \gamma_2=2.0$ (black triangles).
}
\end{figure}
\par
As mentioned in Section \ref{s:qest}, the Helstrom bound to the single-shot 
error probability by quantum probes is bounded from below by the value
$P_e\geq\frac14$, making these kind discrimination schemes of little 
interest for applications. We are thus naturally led to consider
multiple-copy discrimination. In Fig. \ref{f:f4} we report the results
of the optimization of the Chernoff bound of Eq. (\ref{Q}) over 
the parameter $s$ and the interaction time. In the left panel
we show the quantity $Q(\gamma_1,\gamma_2,\lambda)$, minimized over 
the interaction time, as a function of the coupling with the 
environment for different pairs of values $\gamma_1$ and $\gamma_2$
not too close to the limiting values $\gamma=1$ and $\gamma=2$.
Also in this case, the plot also confirms that better performances 
are obtained in the regimes of weak and strong coupling, whereas for 
intermediate values no measurements are able to effectively extract 
information from the quantum probe. The threshold to define the two
regimes increases with the value of the $\gamma$'s themselves.
When the values of the Hurst parameter are approaching the limiting 
values $\gamma=1$ and $\gamma=2$ no threshold appears. In these two 
limiting cases discrimination may be reliably performed in the weak 
coupling limit (for negligible fractal dimension) or in the strong
coupling one (fractal dimension closer to its maximum value).
This behavior is illustrated in the right panel of Fig. \ref{f:f4},
where we show the minimized $Q(\gamma_1,\gamma_2,\lambda)$ 
as a function of the coupling for pairs of values $\gamma_1$ and $\gamma_2$
close to $\gamma=1$ or $\gamma=2$. 
\par
For both, single- and multiple-copy discrimination, the behavior of the optimal
interaction time is analogue to that observed in the discussion of
estimation problem.
\section{Conclusions}\label{s:out}
We have addressed estimation and discrimination problems involving the
fractal dimension of  fractional Brownian noise.  Upon assuming that the 
noise induces a dephasing dynamics on a qubit, we have analyzed in details
the performances of inferences strategies based on quantum limited
measurements.  In particular, in order to assess the
performances of quantum probes, we have evaluated the Bures metric,  the
Helstrom bound and the Chernoff bound, and have optimized their values over
the interaction time.  
\par
Our results show that quantum probes provide an
effective mean to characterize fractional process in two complementary
regimes: Either when the the system-environment coupling is weak,
provided that a long interaction time is achievable, or when the
coupling is strong and the quantum probe may be observed shortly after
that the interaction has been switched on. 
The two regimes of weak and strong coupling are defined in terms of a
threshold value of the coupling, which itself increases with the fractional 
dimension. Our results overall indicate that quantum probes
may represent a valid alternative to characterize
classical noise.
\section*{Acknowledgements}
This work is dedicated to the memory of R. F. Antoni.
The author acknowledges support by MIUR project FIRB LiCHIS-RBFR10YQ3H).

\end{document}